\documentclass{elsart}
\usepackage{natbib}
\usepackage{graphicx}
\usepackage{amssymb}

\begin{document}

\begin{frontmatter}

\title{Orbital phase spectroscopy of X-ray pulsars to study the 
stellar wind of the companion}

\author{U.Mukherjee \& B.Paul}

\address{Department of Astronomy \& Astrophysics,
              Tata Institute of Fundamental Research,
              Homi Bhabha Road, Colaba, Mumbai--400 005,
              India}
\ead{uddipan,bpaul@tifr.res.in}

\begin{abstract}

High Mass X-ray Binary Pulsars (HMXBP), in which the companion star is a 
source of supersonic stellar wind, provide a laboratory to probe 
the velocity and density profile of such winds. Here, we have measured 
the variation of the absorption column density along with other spectral 
parameters over the binary orbit for two
HMXBP in elliptical orbits, as observed with the Rossi X-ray Timing 
Explorer (RXTE) and the BeppoSAX satellites. A spherically symmetric 
wind profile was used as a model to compare the observed column density 
variations. In 4U 1538-52, we find the model corroborating 
the observations; whereas in GX 301-2, the stellar wind appears to be very 
clumpy and a smooth symmetric wind model seems to be inadequate 
in explaining the variation in column density.

\end{abstract}

\begin{keyword}

pulsars : individual (GX~301-2 \& 4U~1538-52) ---
             stars : stellar wind ---
             X-rays : stars

\end{keyword}

\end{frontmatter}

4U~1538-52 (Giacconi et al. 1974) and GX~301-2 (White et al. 1976)
are both luminous ($\sim$10$^{35}$-10$^{37}$ ergs s$^{-1}$) pulsars 
in elliptical orbits. They emit X-rays 
due to the accreted matter from the stellar wind of their respective companion
stars (Parkes et al. 1978, Kaper et al. 1995). 
Since both have well defined orbital measurements (Clark 2000, Koh et al. 1997),
it enables us to inspect the variation of the X-ray spectral parameters 
at different orbital phases to gain useful insight into the morphology of the stellar wind.
In this work, we have measured the X-ray spectral evolution of these two
HMXBP as observed with the RXTE and BeppoSAX satellites and the parameters of the 
spectral model were examined along the orbit 
for any possible variations. We have also compared the equivalent hydrogen 
column density (N$_{\mathrm H}$) variations over the orbit with a model 
variation, assuming a spherically symmetric stellar wind emanating from the companion star.

We observed 4U~1538-52 with RXTE from 2003-07-31 to 2003-08-07
covering out of eclipse phases for two binary orbits. We also
used the archival data from BeppoSAX obtained between 1998-07-29
to 1998-08-01, covering one binary orbit.
GX~301-2 was observed by RXTE, first from 1996-05-10 to 1996-06-15
and again from 2000-10-12 to 2000-11-19.
For more details, refer to Mukherjee and Paul (2004) and Mukherjee et al. (2005) 

For the RXTE data, we took Standard 2 data products of the Proportional 
Counter Array (PCA) and extracted the source spectra 
using the tool saextrct v 4.2d. The BeppoSAX data products 
were extracted from the Medium Energy Concentrator Spectrometers (MECS) and
Low Energy Concentrator Spectrometers (LECS) using circular regions of radii
4$'$ and 8$'$ respectively. The background subtracted source spectra
were analyzed with XSPEC v 11.2.0
(Shafer, Haberl $\&$ Arnaud 1989) using appropriate spectral models.
For 4U~1538-52, we used a power law along with a line-of-sight absorption and a
gaussian line with centre energy $\sim$ 6.4 keV. We also applied a high energy 
cut-off component. 
And for GX~301-2, we used the Partial Covering Absorber Model (PCAM, Endo et al. 2002)
along with the addition of a high energy exponential cutoff. The PCAM is described 
as two different power law components with the same photon index but different 
normalizations (Norm1 \& Norm2); being absorbed by different column densities, 
N$_{\mathrm H1}$ \& N$_{\mathrm H2}$ respectively. $N_{\mathrm H2}$ is interpreted 
as the column density of the
material local to the X-ray source, while $N_{\mathrm H1}$ accounts for the rest
of the material (along with galactic absorption).
In this case, iron K$\alpha$ and K$\beta$ lines and 
an absorption edge at 7.1 keV due to neutral iron were also included in the model.

\section*{Results \& Discussion}

\begin{figure}
\vskip 3.9 cm
\centering
\includegraphics{fig1a.ps}
\includegraphics{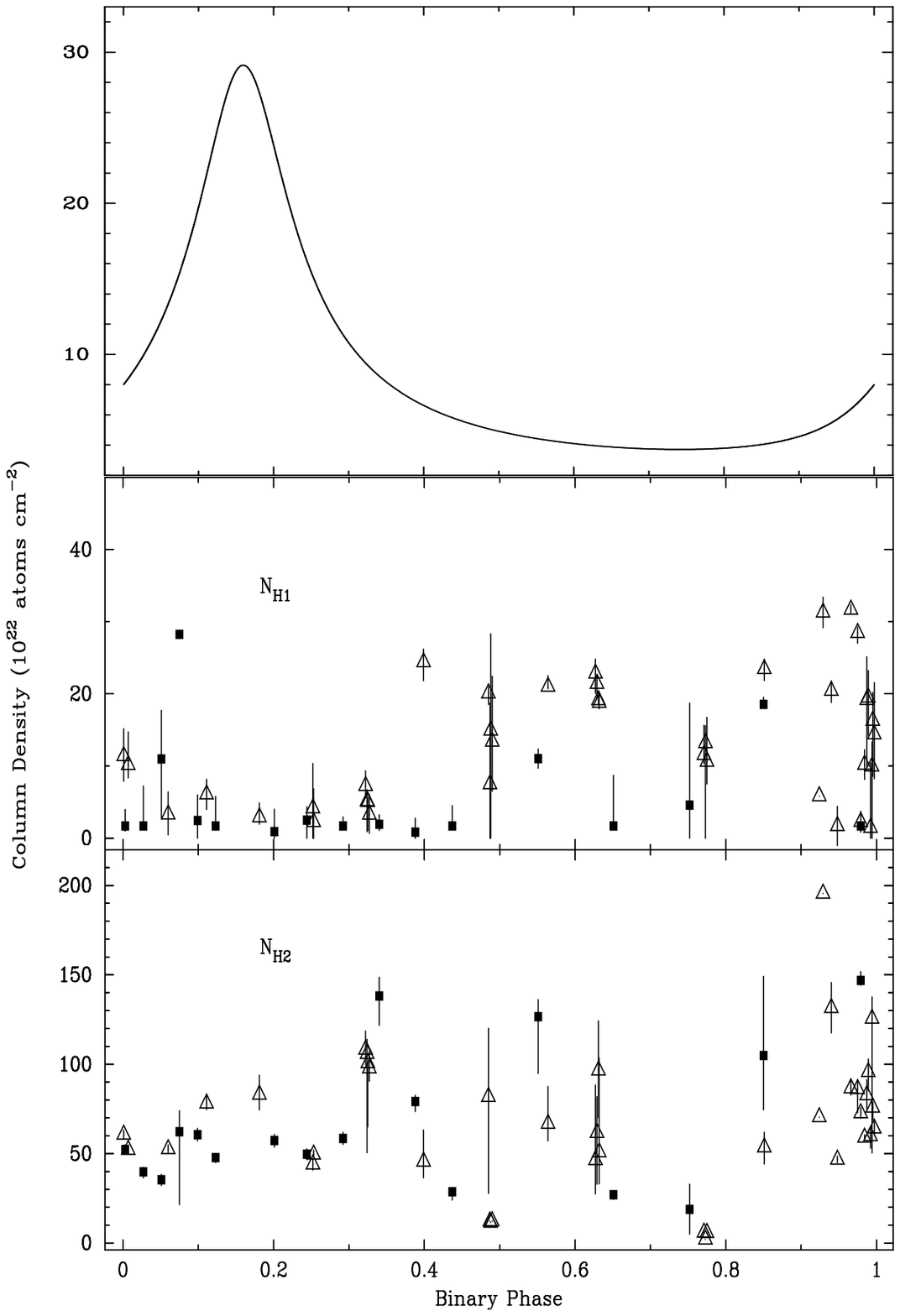}
\caption{The variation of column density (10$^{22}$ H atoms cm$^{-2}$)
with orbital (binary) phase for 4U~1538-52
(left) and GX~301-2 (right) are shown. In the former, the
dashed line, the dashed-dotted and the dotted line represents
the model N$_{\mathrm H}$ values for the inclination angles of 65$^\circ$, 75$^\circ$
\& 85$^\circ$ respectively.
In the latter,
the three panels show the model variation due to the wind and the measured
variations of N$_{\mathrm H1}$ and N$_{\mathrm H2}$ respectively.
In both the figures, different symbols are used for different observations.}
\end{figure}

{\bf 4U~1538-52} : The photon index, fluorescent iron-line
flux, cut-off energy and the e-folding energy measured with 
the average spectrum taken over 2--3 ks
does not show any substantial variation along the orbit. 
This suggests that
the continuum X-ray spectrum of the pulsar is hardly affected during its
revolution along the orbit. But we detected a
notably smooth variation in N$_{\mathrm H}$ with orbital phase, increasing
gradually by an order of magnitude as the pulsar
approaches eclipse (near phases 0.3 and 0.5; as measured from the
periastron passage, Fig. 1). At orbital phases far from the eclipse, the
column density has a value of $\sim$ 1.5 $\times$ 10$^{22}$ H atoms cm$^{-2}$.
We compare the observed column density profile with a model estimated by
assuming a spherically symmetric Castor, Abbott \& Klein (CAK 1975)
wind from the companion star. The velocity profile of the wind is :
$v_{\mathrm wind}$ = {$v_{\infty}$}$\sqrt{1-\frac {{R_\mathrm c}}{r}}$,
where $v_{\infty}$ : terminal velocity of the
stellar wind, R$_{\mathrm c}$ : radius of the companion
and r : radial distance from the centre of the companion star.
The column density profiles were derived using a numerical integration
along the line of sight from the pulsar to the observer for three
different inclination angles of 65$^\circ$, 75$^\circ$ and 85$^\circ$ respectively.
With a mass-loss rate of $\sim$10$^{-6}$ M$_\odot$ yr$^{-1}$ and
v$_\infty$ $\sim$1000 km s$^{-1}$,
the model calculations of N$_{\mathrm H}$ for different
inclination angles when superposed on the observed values show fairly
reasonable agreement (Fig. 1), indicating that a spherically symmetric stellar
wind from the companion star may produce the observed orbital dependence of
the column density for certain range of the orbital inclination.

{\bf GX~301-2} : In most cases, the spectrum was fitted well with the 
PCAM. The average values of the free parameters measured here over the full
binary orbit; viz. photon index, e-folding energy and cut-off energy
follow the general trend which were earlier measured only in some phases
of the binary period (White et al. 1983, Orlandini et al. 2000).
But the variation of N$_{\mathrm H1}$ \& N$_{\mathrm H2}$ with orbital phase was not smooth. 
The values were very 
high with a large variation throughout the binary orbit (from 10$^{22}$ to
10$^{24}$ atoms cm$^{-2}$), indicating a clumpy nature
of the stellar wind (Fig. 1). It is also seen from Fig. 2 that the covering 
fraction (C.F., defined as  Norm2/[Norm1+Norm2]) remains substantially high almost
throughout the orbit. This means that there is dense and clumpy material present 
all through. As in the case of 4U~1538-52, we compared the measured values of column densities
with a model variation using a CAK 
wind. The model values were of the order of 10$^{22}$ to 10$^{23}$ atoms cm$^{-2}$ (Fig. 1).
The peak between phases 0.1 and 0.2 was expected since the line of sight passes through the
densest parts of the wind during these phases. 
Thus it is clear that the observed variation in column density cannot be explained
by a spherically symmetric CAK wind only, indicating
probably strong inhomogeneties in the wind.
Now, as can be seen from Fig. 2, the two iron lines included in our model show large increase in flux
near periastron (phase $\sim$ 0.9) and a possible small increase near phase 0.1 (at least for 1996, filled squares).
The peak near periastron is not very evident in the
1996 data due to the lack of enough observations, though an increasing trend
can possibly be inferred. Moreover, the equivalent width of the 6.4 keV iron-line 
showed a correlation with $N_{\mathrm H2}$, suggesting that most of it is 
produced by the local clumpy matter surrounding the neutron star.

\begin{figure}
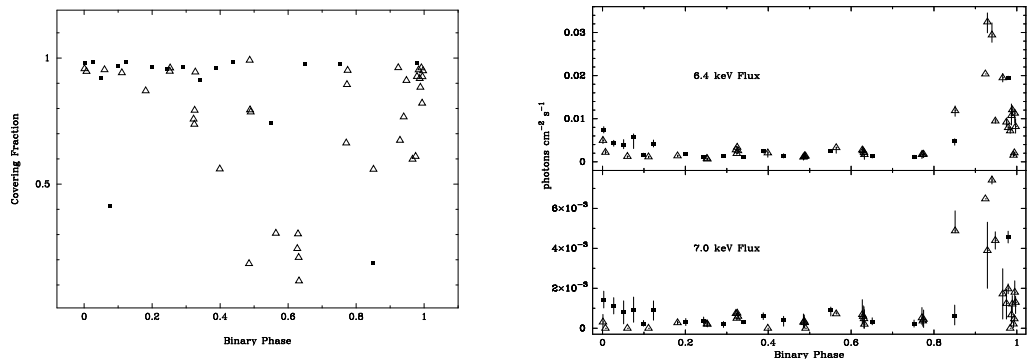

\vskip 2.5 cm
\includegraphics{fig2a.ps}
\includegraphics{fig2b.ps}
\caption{The variation of C.F. (left) and the iron-line flux (right) 
with the binary phase for GX~301-2 are shown with different symbols for 
different observations.}
\end{figure}


\begin{thebibliography}{}


\bibitem [Castor et al.(1975)]{cas} Castor, J.~I., Abbott, D.~C., \& Klein, R.~I., Radiation-driven winds in Of stars, 1975, ApJ, {\bf 195}, p157-174

\bibitem [Clark (2000)]{clark} Clark, G.~W., The Orbit of the Binary X-Ray Pulsar 4U~1538-52 from Rossi X-Ray Timing Explorer Observations, 2000, ApJ, {\bf 542}, L 131-133

\bibitem [Endo et al.(2002)]{endo} Endo, T., Ishida, M., Masai, K., et al., Broadening of Nearly Neutral Iron Emission Line of GX 301-2 Observed with ASCA, 2002, ApJ, {\bf 74}, p879-898

\bibitem[Giacconi et al.(1974)]{gia}
	 Giacconi, R., Murray, S., Gursky, H., et al., The Third UHURU Catalog of X-Ray Sources, 1974, ApJS, {\bf 27}, 
                                                                                                             p37-64

\bibitem [Kaper et al.(1995)]{kap} Kaper, L., Lamers, H. J. G. L. M., Ruymaekers, E., et al., Wray 977 (GX 301-2): a hypergiant with pulsar companion, 1995, A\&A, {\bf 300}, p446-452

\bibitem [Koh et al.(1997)]{koh}
 Koh, D. T., Bildsten, L., Chakrabarty, D., et al., Rapid Spin-up Episodes in the Wind-fed Accreting Pulsar GX~301-2, 1997, ApJ, {\bf 479}, p933-947

\bibitem [Mukherjee \& Paul(2004)]{mukherjee} Mukherjee, U., \& Paul, B., Orbital phase spectroscopy of GX 301-2 with RXTE-PCA, 2004, A\&A, {\bf 427}, p567-573

\bibitem [Mukherjee et al.(2005)]{mukherjee} Mukherjee, U., Raichur, H., Paul, B., et al., Orbital evolution and orbital phase resolved spectroscopy of the HMXB pulsar 4U~1538--52 with RXTE-PCA and BeppoSAX, 2005 (Submitted to ApJ)

\bibitem [Orlandini et al.(2000)]{orl} Orlandini, M., dal Fiume, D., Frontera, F., et al., BeppoSAX Observations of an Orbital Cycle of the X-ray Binary Pulsar GX~301-2, 2000, Adv. Space Res., {\bf 25}, p417-420

\bibitem [Parkes et al.(1978)]{park}
 Parkes, G.~E., Murdin, P.~G., \& Mason, K.~O., The optical counterpart of the binary X-ray pulsar 4U~1538-52, 1978, MNRAS, {\bf 184}, p73-77

\bibitem [Shafer et al.(1989)]{arnaud}
 Shafer, R. A., Haberl, F., \& Arnaud, K. A., XSPEC: An X-ray Spectral Fitting Package, ESA
             TM-09 (Paris:ESA), 1989

\bibitem [White et al.(1976)]{white} White, N. E., Mason, K. O., Huckle, H. E., et al., Periodic modulation of three galactic X-ray sources, 1976, ApJ, {\bf 209}, L 119-124

\bibitem [White et al.(1983)]{white2} White, N. E., Swank, J. H., \& Holt, S. S., Accretion powered X-ray pulsars, 1983, ApJ, {\bf 270}, p711-734

\end{thebibliography}
\end{document}